\documentclass[12pt]{article}
\usepackage{epsfig}
\usepackage{graphicx}

\newcommand{\be}{\begin{equation}}
 \newcommand{\ee}{\end{equation}}
 \newcommand{\bea}{\begin{eqnarray}}
 \newcommand{\eea}{\end{eqnarray}}
 \newcommand{\nn}{\nonumber}

\usepackage{latexsym}
 \usepackage{amssymb}

 \date{October 26, 2005}

\title{ BASICS OF A PHOTON COLLIDER\footnote{Lectures
for young physicists in the series of dedicated lectures on Physics
at Future Colliders (LHC, ILC, PLC) and Astrophysics at the
International Conference ``The PHOTON: its first hundred years and
the future'' (30.08 --- 08.09.2005, Warsaw and Kazimierz, Poland).
This work is supported in part by RFBR (code 03-02-17734) and by the
Fund of Russian Scientific Schools (code 2339.2003.2). The author is
grateful to the Organizing Committee for warm hospitality and
support. }}
\author{Valery G.~SERBO\\
 {\it  Novosibirsk State University, 630090 Novosibirsk, Russia}
}

\begin{document}

\maketitle

\begin{abstract}

This small review is devoted to $\gamma\gamma$ collisions including
methods of creating the colliding $\gamma \gamma$ beams of high
energy and physical problems which can be solved or clarified in
such collisions.\\
Contents:

1. Introduction

\hspace{3mm} {\it  1.1. The subject}

\hspace{3mm} {\it 1.2. Interaction of photons in the Maxwell theory
and QED}

2. Collisions of equivalent photons at $e^+e^-$ storage rings

3. Results obtained in virtual $\gamma^*\gamma^*$ collisions

4. Linear $e^{\pm} e^-$ collider

5. Photon collider on a base of a linear $e^{\pm} e^-$ collider

\hspace{3mm} {\it 5.1. Idea of high-energy $\gamma\gamma$ and
$\gamma e$ colliders with real photons}

\hspace{3mm} {\it 5.2. Scheme of a photon collider}

\hspace{3mm} {\it 5.3. Compton scattering as a basic process for
$e\to \gamma$ conversion}

6. Physics of $\gamma\gamma$ interactions

7. Concluding remarks

\hspace{3mm} {\it 7.1. Summary from TESLA TDR}

\hspace{3mm} {\it 7.2. Prediction of Andrew Sessler in 1998}

\hspace{3mm} {\it 7.3. Conclusion of Karl von Weizs\"acker for young
physicists}
\end{abstract}

\section{Introduction}

\subsection{The subject}

This small review is based on papers \cite{BGMS-75},
\cite{GKST-Springer}, \cite{TESLA-04}. It deals with high-energy
$\gamma\gamma$ collisions, this is a new and promising area in
high-energy physics related to fundamental problems of strong and
electro-weak interactions.

Our knowledge about elementary particles and their interactions is
mainly obtained from particle collisions. Accelerators with
colliding beams are called now {\it colliders}. Most of fundamental
results in particle physics have been obtained from experiments at
the $pp,\; p\bar p, \;e^+e^-$ and $ep$ colliders.

Principal characteristics of colliders are:
\begin{enumerate}
\item{the {\it energy} in the center-of-mass system (c.m.s)
$E_{\rm cm}$;}
\item{{\it luminosity} of a collider $L$ which determines
collision rate $\dot N$ of events with the cross section $\sigma$ by
relation $\dot N = L\, \sigma$;}
\item{{\it types} of colliding particles.}
\end{enumerate}

The progress on high-energy colliders can be seen from Table
\ref{T:1}. Up to now and in the nearest future, the $pp$ and $ p\bar
p$ colliders are the machines with the highest energy. That is why
such epochal discoveries as $W$ and $Z$ bosons (responsible for weak
interaction) and $t$ quark had been performed at the S$p\bar p$S and
the TEVATRON, respectively. It is not excluded that at future $pp$
collider LHC the Higgs boson $ H$ (thought to be responsible for the
origin of the particle masses) will be discovered.

\begin{table}[th]
\centering
 \caption{High energy colliders}
\begin{tabular}{lccl}\hline
Collider & Type & $ E_{\rm cm} $, TeV & Start date \\  \hline
&&& \\[-2mm]
S$p\bar p$S & $p\bar p$ & 0.6 & 1981 \\
TEVATRON & $p\bar p$ & 2 & 1987 \\
LHC & $ p p$ & 14 & 2007 \\[2mm]
HERA & $e p$ & 0.31 & 1992 \\[2mm]
SLC & $e^+e^-$ & 0.1 & 1989 \\
LEP-I & $e^+e^-$ & 0.1 & 1989 \\
LEP-II &$ e^+e^- $ & 0.2 & 1999 \\[2mm]
Linear collider & $e^+e^-$ & 0.5 & 2010? \\
Photon collider & $\gamma \gamma,\; \gamma e$ & 0.4 & 2010+?
\\[2mm]
Muon collider & $\mu^+\mu^-$ & 0.1$\div$ 3 & ?? \\
\end{tabular}
 \label{T:1}
\end{table}

For detail study of new phenomena, it is important not only the
energy but also types of colliding particles. The $e^+e^-$
colliders, being less energetic then $pp$ colliders, have some
advantages over proton colliders due to much lower background and
simpler initial state. Well known example --- the study of $Z$
boson. It was discovered at $ p\bar p$ collider S$p\bar p$S where
about 100 events with $Z$ bosons were found among about $10^{11}$
background events, while the detailed study of $Z$ boson had been
performed at the $e^+e^-$ colliders LEP and SLC which provided us
with more than $10^7$ $Z$-events with a very low background.

About thirty years ago a new field of particle physics ---
photon-photon interactions --- has
appeared~\cite{Bal71}--\cite{BKT}. Up to now $\gamma \gamma$
interactions were studied in collisions of virtual (or equivalent)
photons at $e^+e^-$ storage rings. It was obtained a lot of
interesting results. However, the number of the equivalent photons
is by 2 order of magnitude less than the number of electrons.

A new possibility in this field is connected with high-energy
$e^{\pm}e^-$ linear colliders which are now under development. The
electron bunches in these colliders are used only ones. This makes
possible to ``convert'' electrons to real high-energy photons, using
the Compton back-scattering of laser light, and thus to obtain
$\gamma \gamma$ and $\gamma e$ colliders with real
photons~\cite{GKST81a}--\cite{GKPST84}. The luminosity and energy of
such colliders will be comparable to those of the basic $e^{\pm}e^-$
colliders.

Physical problems, which are now investigated in the $\gamma \gamma$
collisions, are mainly connected with strong interaction at large
$\sim \hbar /(m_\pi c)$ and moderate small distances $\sim \hbar
/p_\bot$, where $p_\bot \; \stackrel{<}{\sim} 10$ GeV/c. In future
it will be a continuation of the present day experiments plus
physics of gauge $(W^\pm,\; Z)$ bosons and Higgs $H$ bosons, i.e. it
will be problems of the electro-weak interactions, Standard Model
and beyond, search of new particles and new interactions. In other
words, physics at high-energy $\gamma \gamma$ and $\gamma e$
collisions will be very rich and no less interesting than at $pp$ or
$e^+e^-$ collisions. Moreover, some phenomena can best be studied at
photon collisions.

At the end of this subsection it will be appropriate to cite some
words from the article ``Gamma-Ray Colliders and Muon Colliders'' of
the former President of the American Physical Society A.~Sessler in
{\it Physics Today}~\cite{Sessler}:

\begin{quotation}

In high-energy physics, almost all of the present accelerators are
colliding-beam machines.

In recent decades these colliders have produced epochal discoveries:
Stanford SPEAR electron-positron collider unveiled the charmed-quark
meson and $\tau$ lepton in 1970s. In the realm of high-energy
proton-antiproton colliders, the Super Proton Synchrotron at CERN
gave us the W$^{\pm}$ and Z$^0$ vector bosons of electroweak
unification in 1990s, and in 1999s the Tevatron at Fermilab finally
unearthed the top quark, which is almost 200 times heavier than the
proton.

...What about other particles? Beam physicists are now actively
studying schemes for colliding photons with one another and schemes
for colliding a beam of short-lived $\mu^+$ leptons with a beam of
their $\mu^-$ antiparticles.

If such schemes can be realized, they will provide extraordinary new
opportunities for the investigation of high-energy phenomena.

These exotic collider ideas first put forward in Russia more that 20
years ago...

\end{quotation}

\begin{figure}[!h]
\centering
\includegraphics[width=8cm,angle=0]{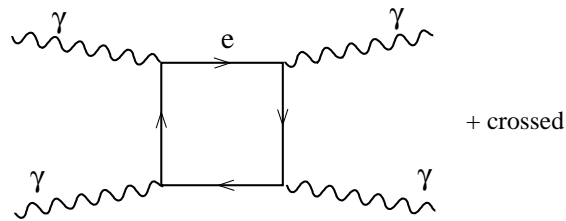}
\caption{
Feynman diagrams for the elastic $\gamma \gamma$  scattering in QED}
 \label{F1.1}
\end{figure}

\subsection{Interaction of photons in the Maxwell theory and QED }

\begin{figure}[!htb]
\centering
\includegraphics[width=9cm,angle=0]{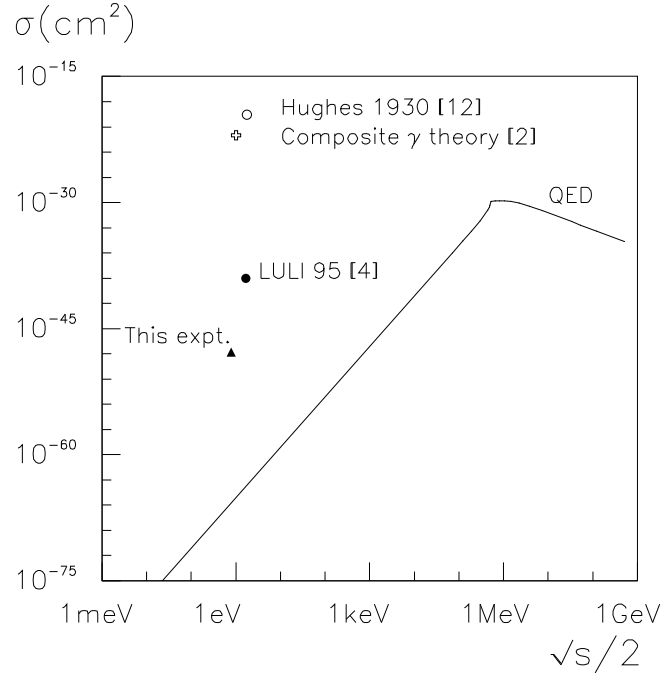}
\caption{Elastic photon cross section as a function of photon c.m.s.
energy}
 \label{cro}
\end{figure}

The Maxwell's equations are linear in the strengths of the electric
and magnetic fields. As a result, in the classical Maxwell theory of
electromagnetism, rays of light do not interact with each other. In
quantum electrodynamics (QED) photons can interact via virtual
$e^+e^-$ pairs. For example, an elastic $\gamma \gamma$ scattering
is described by Feynman diagrams of Fig.~\ref{F1.1}. The maximal
value of the cross section is achieved at the c.m.s. photon energy
$\omega \sim m_ec^2$ and is large enough:
\begin{equation}
\max \;\sigma_{\gamma \gamma \to \gamma \gamma} \sim \alpha^4
\left({\hbar\over m_ec}\right)^2= \alpha^2 r^2_e\sim 4\cdot
10^{-30}\mbox{ cm}^2. \label{1.3}
\end{equation}
However, at low energies, $\omega \ll m_ec^2$, this cross section is
very small
\begin{equation}
\sigma_{\gamma \gamma \to \gamma \gamma}=0.031 \alpha^2 r^2_e
\left({\omega\over m_ec^2}\right)^6\,. \label{1.1}
\end{equation}
For example, for visible light, $\omega \sim 1$ eV,
\begin{equation}
\sigma_{\gamma \gamma \to \gamma \gamma} \sim 10^{-65} \; \mbox{cm}
^2\,. \label{1.2}
\end{equation}
It is too small to be measured even with the most powerful modern
lasers, though there were such attempts. In recent
paper~\cite{Bernard} it was obtained an upper limit of the cross
section of $\sigma(\gamma \gamma \to \gamma \gamma)_{\rm
Lim}=1.5\times 10^{-48}$ cm$^2$  for the photon c.m.s. energy $0.8$
eV (see Fig.~\ref{cro} from ~\cite{Bernard}).

\begin{figure}[!htb]
\centering
\includegraphics[width=8cm,angle=0]{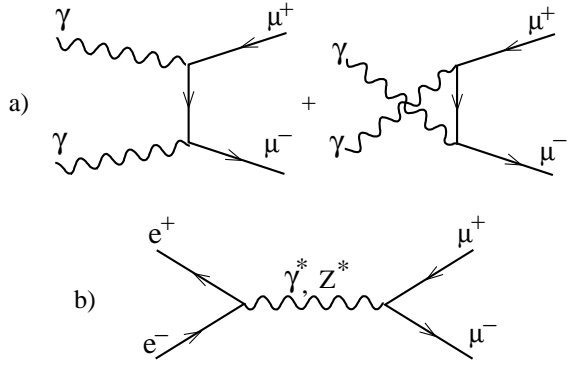}
\caption{
Feynman diagrams for a) $\gamma \gamma \to \mu^+\mu^-$ and b)
$e^+e^- \to \mu^+\mu^-$} \label{F1.2}
\end{figure}

At energies $\omega > m c^2$, two photons can produce a pair of
charged particles. The cross section of the characteristic process
$\gamma \gamma \to \mu^+ \mu^-$ (Fig.\ref{F1.2}a) is equal to
\begin{equation}
\sigma_{\gamma \gamma \to \mu^+ \mu^-} = 4\pi r_e^2
\frac{m_{e}^2c^4}{s} \, \ln{s\over m^2_\mu c^4} \;\;\;\; {\rm at}
\;\;\;\; s=(2\omega)^2 \gg 4m^2_\mu c^4\,.
 \label{1.5}
\end{equation}
It is larger than the ``standard'' cross section for the production
of the same pair in the $e^+ e^-$ collisions (Fig.\ref{F1.2}b via a
virtual photon only)
\begin{equation}
\sigma_{e^+e^- \to \mu^+ \mu^-} = \frac{4}{3}\,\pi r_e^2\,
\frac{m_{e}^2c^4}{s}\,.
 \label{1.5a}
\end{equation}

\subsection{Collisions of equivalent photons at $e^+ e^-$
storage rings}

Unfortunately, there are no sources of intense high-energy photon
beams (like lasers at low energies). However, there is indirect way
to get such beams --- to use equivalent photons which accompanied
fast charged particles. Namely this methods was used during last
three decades for investigation of two-photon physics at $e^+e^-$
storage rings. The essence of the equivalent photon approach can be
explained in the following way~\cite{Fermi,WW} (see also~\cite{BLP}
\S 99). The electromagnetic field of an ultra-relativistic electron
is similar to the field of a light wave. Therefore, this field can
be described as a flux of the {\it equivalent} photons with energy
distribution $dn_\gamma / d\omega$. The number of these photons per
one electron with the energy $E$ is
\begin{equation}
dn_\gamma\sim {2\alpha\over \pi}\;\ln{E\over \omega}\,
{d\omega\over\omega} \label{1.6}
\end{equation}
or approximately
\begin{equation}
dn_\gamma\sim 0.03\;{d\omega\over\omega}\,.
 \label{1.7}
\end{equation}
At the $e^+e^-$ colliders the equivalent photons also collide and
can produce some system of
\begin{figure}[!htb]
\centering
\includegraphics[width=11cm,angle=0]{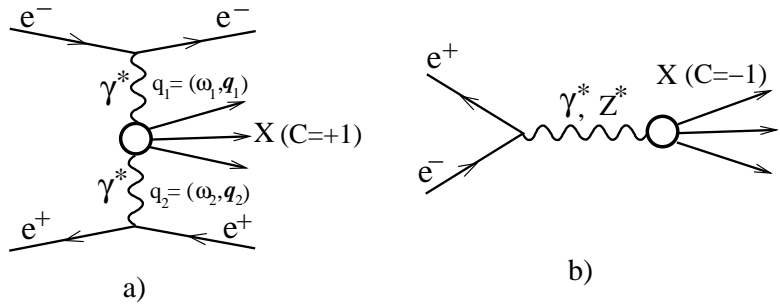}
\caption{
Production of system $X$ a) by two equivalent photons $\gamma^*$
with 4-momenta (energies) $q_1\;(\omega_1)$ and $q_2\; (\omega_2)$
emitted by an electron and a positron and b) in the annihilation
process $e^+e^- \to X$ } \label{F1.3}
\end{figure}
particles $X$ (see Fig.~\ref{F1.3}a, $\gamma^*$ denotes the
equivalent photon)
\begin{equation}
e^+ e^-\to e^+ e^-\gamma^\ast\gamma^\ast\to e^+ e^- X\,. \label{1.8}
\end{equation}
Thus, this process is directly connected with the subprocess
$\gamma^* \gamma^* \to X$. Strictly speaking, the equivalent photons
are not real photons, they are virtual ones. The 4-momentum squared
of such a photon $q_i^2$ (which is equal to $m^2 c^2$ for usual
particle) is not equal zero, $q_i^2 \neq 0$. But for large part of
the cross section $|q^2_i|$ is very small, therefore, the most of
equivalent photons are almost real.

The cross section for two-photon production of  $e^+ e^-$ in
collisions of two fast particles with charges $Z_1e$ and $Z_2e$,
i.e. for the $Z_1 Z_2 \to Z_1 Z_2 e^+ e^-$ process, was calculated
by Landau and Lifshitz \cite{LL} in 1934 (see also \cite{BLP} \S
100). In fact, it was the PhD of E.M.~Lifshitz.

At first sight, the cross sections of the two-photon processes at
$e^+ e^-$ colliders (Fig.~\ref{F1.3}a) are very small since they are
the 4-order processes: $ \sigma_{\rm two-phot} \propto \alpha^4\,, $
while for the annihilation processes of Fig.~\ref{F1.3}b the cross
sections $ \sigma_{\rm annih} \propto \alpha ^2. $ However, the
annihilation cross sections decrease with increase of the energy
(compare with (\ref{1.5a}))
\begin{equation}
\sigma_{\rm annih}\sim{\alpha^2}\,{\hbar^2 c^2\over s},\quad
s=(2E)^2,
 \label{1.9}
\end{equation}
while the two-photon cross sections increase
\begin{equation}
\sigma_{\rm two-phot}\sim{\alpha^4}\,{\hbar^2\over m^2_{\rm
char}c^2}\, \ln^n{s}\,.
 \label{1.10}
\end{equation}
Here $n=3\div 4$ depending on the process, and the characteristic
mass $ m_{\rm char}$ is constant (for example, $ m_{\rm char}\sim
m_\mu$ for $X= \mu^+ \mu^-$ and $ m_{\rm char}\sim m_\pi$ for $X=
hadrons$). As a result, already at $\sqrt{s}> 2$ GeV
\begin{equation}
\sigma_{e^+ e^-\to e^+ e^-\mu^+\mu^-} > \sigma_{e^+
e^-\to\mu^+\mu^-}. \label{1.11}
\end{equation}
Another  example, at the LEP-II electron-positron
\begin{figure}[htb]
\centering
\hspace*{5mm}\includegraphics[width=11cm,angle=0]{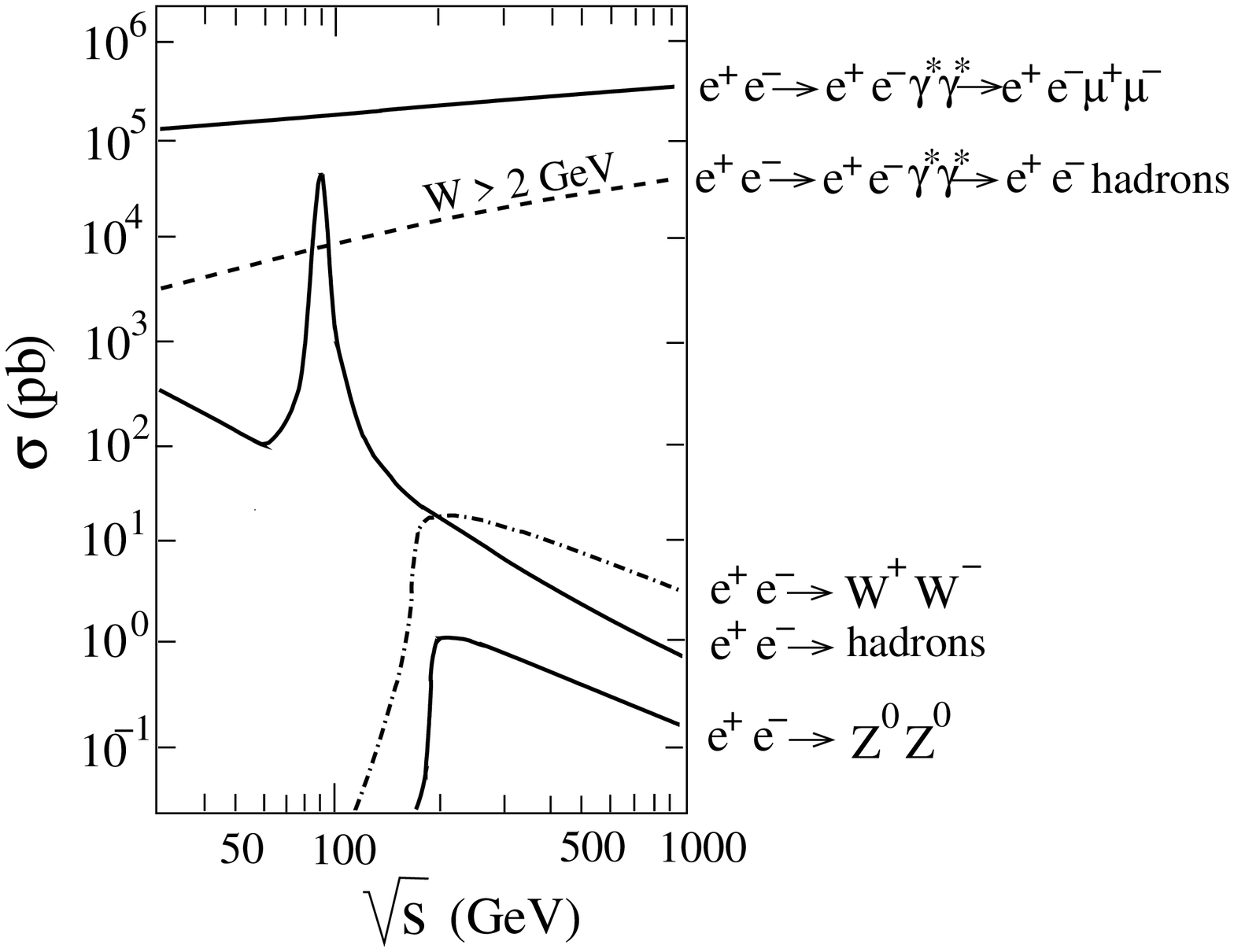}
\caption{
Cross sections for some annihilation and two-photon processes in
$e^+e^-$ collisions}
 \label{F1.4}
\end{figure}
collider with the energy $\sqrt{s}= 200$ GeV, the number of events
for two-photon production of hadrons with the c.m.s. energy
$W_{\gamma \gamma} > 2$ GeV is by a three order of magnitude larger
than that in the annihilation channel (Fig.\ref{F1.4}).

At $e^+e^-$ storage rings the first two-photon processes $e^+e^- \to
e^+e^- e^+e^-$ had been observed in 1970 (Novosibirsk~\cite{Bal71}).
The importance of two-photon processes for the lepton and hadron
production at $e^+e^-$ storage rings had been emphasized in the
papers Arteage-Romero, Jaccarini, Kessler and Parisi~\cite{Kes},
Balakin, Budnev and Ginzburg~\cite{BBG} and Brodsky, Kinoshita and
Terazawa~\cite{BKT}. In the papers~\cite{BBG} it  was shown that
$e^+e^-$ colliding beam experiments can give information about a new
fundamental process $\gamma^* \gamma^* \to hadrons$ and the
necessary formulae and estimations were obtained.

At that time there were a lot of theoretical investigations of
various aspects of two-photon physics , but only a few experimental
results~\cite{Bal71,Fras} have been obtained related mainly to the
processes $\gamma\gamma \to e^+ e^-$, $\gamma\gamma \to\mu^+ \mu^-$.
This period of two-photon physics was summarized in review by
Budnev, Ginzburg, Meledin and Serbo~\cite{BGMS-75}.

A few years later (approximately from 1977) it was shown in a number
of theoretical papers that the two-photon processes are very
convenient for the test and detailed study of the Quantum
Chromodynamics (QCD) including investigation of
\begin{itemize}
\item a photon structure function (Witten \cite{Witten}),
\item  a jet production in the $\gamma \gamma$ collisions
(Llewelyn Smith \cite{LS}; Brodsky, De Grand, Gunion and Weis
\cite{BGGW}; Baier, Kuraev and Fadin \cite{BKF}),
\item  the $\gamma \gamma \to c {\bar c} c {\bar c}$ process and
the problem of the perturbative Pomeron (Balitsky and Lipatov
\cite{BL}).
\end{itemize}

A new wave of experimental activity in this field was initiated by
the experiment at SLAC~\cite{ATel79} which demonstrated that
two-photon processes can be successfully studied without detection
of the scattered electrons and positrons. After that there was a
flow of experimental data from almost all detectors at the $e^+e^-$
storage rings. It should be noted a special detector MD-1 in
Novosibirsk with a transverse magnetic field in the interaction
region and system of registration of scattered at small angles
electrons which was developed for two-photon experiments (see review
\cite{MD1}). This period was reviewed by Kolanoski \cite{Kol84}.
\begin{figure}[!h]
\centering
\includegraphics[width=12cm,angle=0]{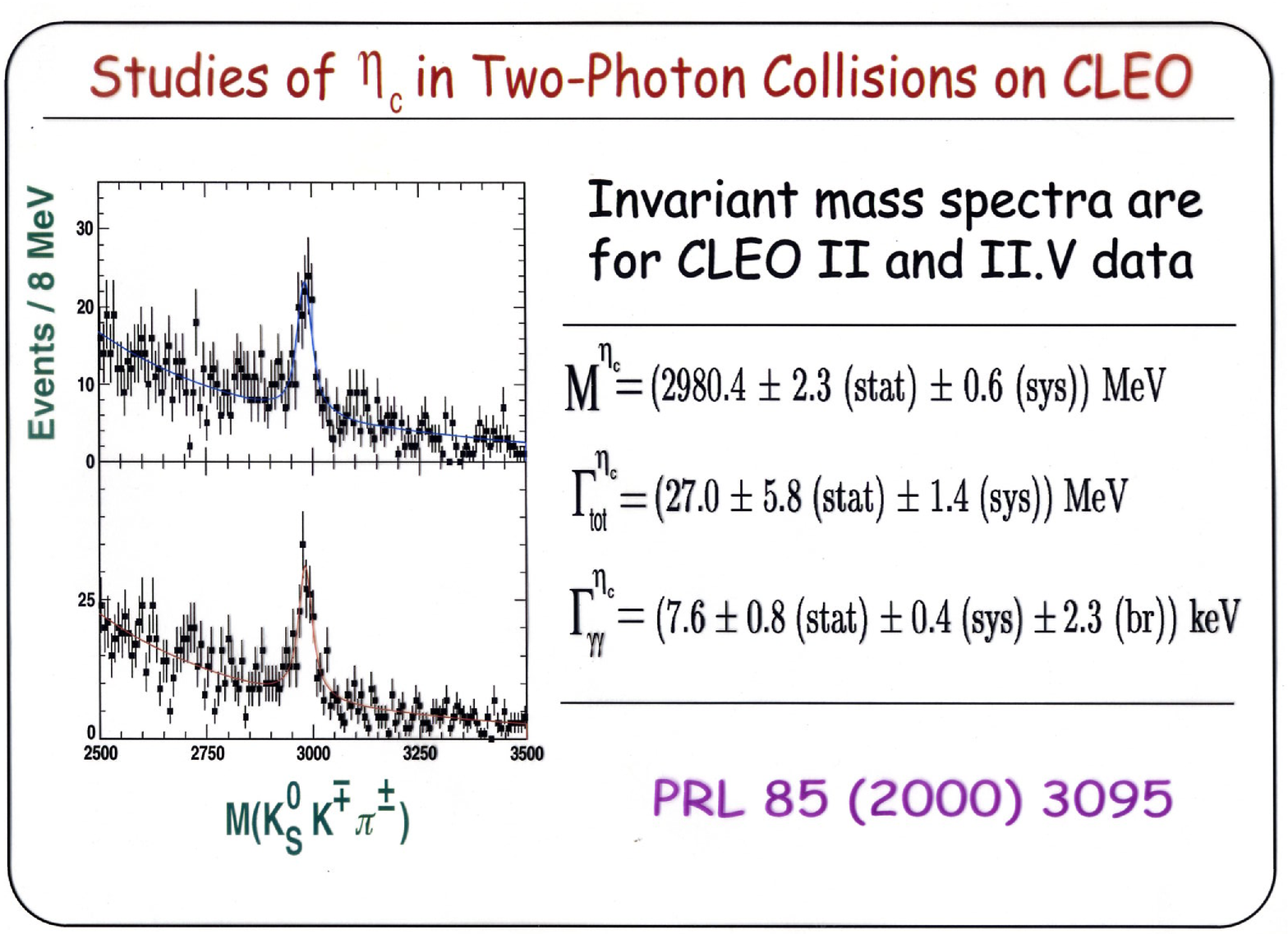}
\caption{Results for the process $\gamma \gamma \to \eta_c$}
\label{savinov}
\end{figure}

\section{Results obtained in virtual $\gamma^* \gamma^* $
collisions}

In experiments at $e^+ e^-$ storage rings a lot of interesting
results about $\gamma^* \gamma^* $ collisions have been obtained
(see reviews \cite{Kol84,Pope,MPW94} and Proceedings of Workshops on
Photon-Photon Collisions), among them:
\begin{itemize}
\item production of $C$-even resonances in $\gamma ^* \gamma ^*
$ collisions, such as $\pi ^0,\; \eta,$ $ \eta ^{\prime}, \; f_2,\;
a_2$, $ \eta_c$, $\chi_c$, ...  and measurement of their $\gamma
\gamma$ width;
\item measurement of the total $\gamma \gamma \to hadrons$
cross section up to c.m.s. energy $W_{\gamma \gamma} $ about 150
GeV;
\item measurement of the total $\gamma ^* \gamma ^* \to hadrons$
cross section with large values of $W^2_{\gamma \gamma} $ and photon
virtualities $-q^2_1 \sim -q^2_2 \sim 10$ (GeV/c)$^2$;
\item a number of exclusive reactions: $\gamma ^* \gamma ^*
\to \pi \pi, \; K \bar K,\; p \bar p,\; \rho \rho,\; \rho \omega,$
etc.;
\item  investigation of the photon structure function in the
collision of almost real photon and highly virtual photon with
$-q^2$ up to about 1000 (GeV/c)$^2$;
\item jet production in $\gamma \gamma$ collisions.
\end{itemize}
As an example, let us present the beautiful result (Fig.
\ref{savinov}) for the study of two-photon process  $\gamma \gamma
\to \eta_c$ at CLEO (Cornell).

Unfortunately, the number of equivalent photons per one electron is
rather small, and correspondingly the $\gamma ^* \gamma ^*$
luminosity is about $3\div 4$ orders of magnitude smaller than that
in $e^+e^-$ collisions. Therefore, it is not surprising that the
most important results at $e^+e^-$ storage rings were obtained in
the $e^+e^-$ annihilation.

\section{Linear $e^{\pm} e^-$ collider }

\begin{figure}[!h]
\begin{center}
\includegraphics[width=14cm,angle=0]{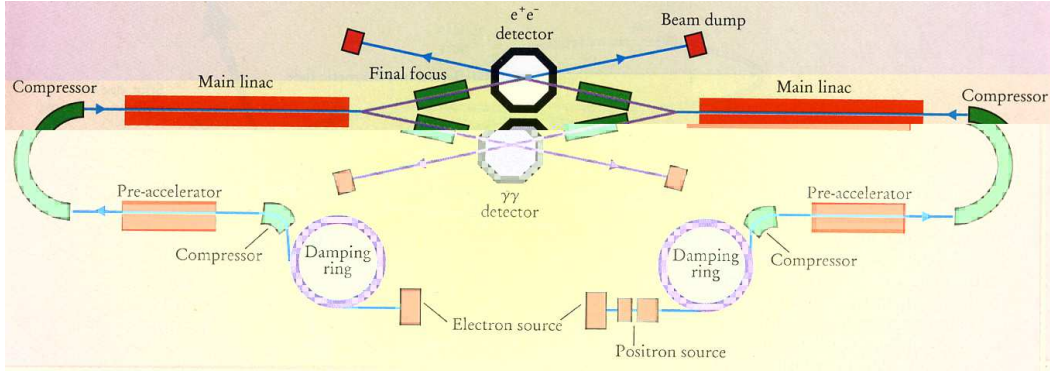}
\end{center}
 \caption{Schema of a linear $e^+e^-$ collider}
  \label{schema-ses-ee}
 \end{figure}

New opportunities for two-photon physics are connected with future
linear $e^{\pm}e^-$ colliders. Projects of such accelerators are now
under development in several laboratories. A linear collider
consists of several main systems (see Fig. \ref{schema-ses-ee}
from~\cite{Sessler}): electron injectors, pre-accelerators, a
positron source, two damping rings, bunch compressors, main linacs,
interaction regions, a beam dump.

Since 1988 this field is developed in a very tight international
collaboration of physicists from many countries. In 1996-97 three
projects NLC (North America), JLC (Asia) and TESLA (Europe) have
published their Conceptual Design Reports~\cite{Projects} of the
linear colliders in the energy range from a few hundred GeV to about
one TeV; in 2001 the TESLA Technical Design Report~\cite{TESLA},
\cite{TESLA-04} has been published. One team at CERN is working now
on the conception of multi-TeV Compact Linear Collider (CLIC).
Current parameters of these projects are presented in
Table~\ref{t2.1}. Parameters of the projects NLC and JLC are
presented in one column since their teams developed a common set of
the collider parameters.
\begin{table}[!h]
 \caption{Some parameters of the linear colliders  NLC/JLC and TESLA}
 \vspace{.2cm}
\renewcommand{\arraystretch}{1} \setlength{\tabcolsep}{1.1mm}
\begin{center}
\begin{tabular}{lccc} \hline &&NLC/JLC&TESLA
\\ \hline \hline
C.m.s. energy $2E_0$& [TeV]& 0.5 & 0.5 \\
Luminosity $L$& [$10^{34}$/(cm$^{2}$s)]&  2.2 & 3 \\
Repetition rate $f_r$ &[Hz]&  120 & 5 \\
No. bunch/train $n_b$&  & 190& 2820\\
No. particles/bunch $N_e$& [$10^{10}$] &  0.75 & 2 \\
Collision rate $\nu$& kHz &  22.8 & 14.1 \\
Bunch spacing $\Delta t_b$& [ns] &  1.4 & 337  \\
Accel. gradient $G$ & [MeV/m]&  50 & $\sim 25$\\
{Linac length} $L_l$ & {[km]}& { 10} &{ 20} \\
Beams power $2P_b$& [MW]& 14 & 22.5 \\
IP beta-function $\beta_x/\beta_y$& [mm] &   8/0.1& 15/0.4 \\
{R.m.s. beam size at IP} $\sigma_x/\sigma_y$ \hspace{-5mm}
& {[nm]}& {245/2.7}& {555/5} \\
R.m.s. beam length $\sigma_z$ & [$\mu$]&  110&300 \\
 \hline
\end{tabular}
\end{center}
\label{t2.1}
\end{table}

Now the project of International Linear Collider (ILC) is under
development.

A few special words have to be said about luminosity of the linear
colliders, which is determined as
  \be
L=\nu\;{N_{e^+}\, N_{e^-}\over S_{\rm eff}}\,,
 \ee
where the effective transverse bunch aria $S_{\rm eff}\sim \sigma_x
\sigma_y$. In the next table we present a comparison of the storage
ring LEP-II and linear colliders:
\begin{center}
\begin{tabular}{|c|c|c|c|c|c|c|}
 \hline
Collider & $ 2E_{e}$, &  $L$, $10^{33}$ & $\nu$, & $N_{e^{\pm}},$ &
$\sigma_x$, &$\sigma_y$, \\ &  GeV &1/(cm$^{2}$s) &
 kHz &  $10^{10}$ &
$\mu$ & nm\\  \hline LEP-II & 200 & 0.05 & 45 & 30 & 200 & 8000\\
\hline NLC/JLC & 500 & 22 & 22.8 & 0.75 & 0.245  & 2.7 \\
\hline TESLA   & 500 & 30  & 14.1   & 2    & 0,555 & 5 \\
\hline
\end{tabular}
\end{center}
Note  transverse bunch sizes
  $$
{\rm LEP-II}: \;\;\;\;\;\;\sigma_x \sigma_y \;\;\sim \;\; 10^{-5}
{\rm cm}^2\,,
  $$
 $$
{\rm TESLA}: \;\; \sigma_x \sigma_y \sim  3\cdot 10^{-11} {\rm
cm}^2\,.
 $$

 So, it is likely that a first linear collider  will have energy
about 500 GeV with some possible extension up to 1.5 TeV. Compared
to the LEP, these colliders of the so called next generation are
designed on 2.5--7 times higher energy and four orders of magnitude
higher luminosity!

\section{Photon collider on a base of a linear
$e^{\pm} e^-$ collider}

\subsection{Idea of high-energy $\gamma\gamma$ and
$\gamma e$ colliders with real photons}

Unlike the situation in storage rings, {\it in linear colliders each
$e^{\pm}$ bunch is used only once}. It makes possible to ``convert''
electrons to high-energy photons and to obtain the $\gamma \gamma$
or $\gamma e$ colliding beams with approximately the same energy and
luminosity\footnote{ This can not be done at usual $e^+e^-$ storage
rings where a high luminosity is provided by large number of
collisions ($\sim 10^9-10^{11} $) of the same $e^+$ and $e^-$
bunches. Conversion of the electron and positron bunches into the
$\gamma $ bunches at the storage ring gives only a single collision
of gamma bunches. The resulting luminosity will be very low because
obtaining of new $e^+$ and $e^-$ bunches at storage rings takes long
time.} as in the basic $e^{\pm}e^-$ collisions. Moreover, $\gamma
\gamma$ luminosity may be even larger due to absence of some
collisions effects.

This idea was put forward by Novosibirsk group in 1981--1983
(Ginzburg, Kotkin, Serbo and Telnov \cite{GKST81a}--\cite{GKST83})
and was further developed in detail.

Among various methods of $e\to \gamma$ conversion (bremsstrahlung,
ondulator radiation, beamstrahlung and so on), the best one is the
Compton scattering of laser light on high-energy electrons. In this
method a laser photon is scattered backward taking from the
high-energy electron a large fraction of its energy. The scattered
photon travels along the direction of the initial electron with an
additional angular spread $\theta \sim 1/\gamma_e$ where $\gamma_e
=E/(m_e c^2)$ is the Lorentz factor of the electron. This method was
known long time ago~\cite{ATM63} and has been realized in a number
of experiments, e.g.~\cite{FIAN64,SLAC69}. However, the conversion
coefficient of electrons to high-energy photons $k=N_\gamma / N_e$
was very small in all these experiments. For example, in the SLAC
experiment \cite{SLAC69} it was $\sim 10^{-7}$.

In our papers \cite{GKST81a,GKST83} it was shown that at future
linear $e^{\pm}e^-$ colliders {\it it will be possible to get}
$k\sim 1$ at a quite reasonable laser flash energy of a few Joules.

Therefore, two principal facts, which make possible a photon
collider, are:

\begin{itemize}

\item linear colliders are single-pass accelerators, the
electron beams are used here only once;

\item obtaining of conversion coefficient $k\sim 1$ is
technically feasible.

\end{itemize}

It should be noted that {\it positrons are not necessary for photon
colliders}, it is sufficient and much easier to use the
electron-electron colliding beams.

The problems of the $\gamma \gamma$ and $\gamma e$ colliders were
discussed on many conferences: Photon-Photon Collisions, Linear
Colliders, and dedicated $\gamma \gamma$ Workshops~\cite{gg94,gg00}.
Very rich physics, potentially higher than in $e^+e^-$ collisions
luminosity, simplification of the collider (positrons are not
required) are all attractive to physicists. Progress in development
of linear $e^+e^-$ colliders and high power lasers (both
conventional and free-electron lasers) makes it possible to consider
photon colliders as very perspective machines for investigation of
elementary particles.

This option has been included in Conceptual~\cite{Projects} and
Technical~\cite{TESLA} Designs of linear colliders. All these
projects foresee the second interaction regions for the $\gamma
\gamma$ and $\gamma e$ collisions.

\subsection{Scheme of a photon collider}

To create a $\gamma \gamma$ or $\gamma e$ collider with parameters
comparable to those in $e^+e^-$ colliders, the following
requirements should be fulfilled:

\begin{itemize}
\item the photon energy $\omega \approx E_0$ from $100$ GeV
to several TeV;
\item the number of high-energy photons
$N_\gamma \sim N_e \sim 10^{10}$;
\item photon beams should be focused on the spot with transverse sizes
close to those which electron bunches would have at the interaction
point $\sigma_x \times \sigma_y \sim 10^{-5}$ cm $\times 10^{-7}$
cm.
\end{itemize}

\begin{figure}[!h]
\centering
\includegraphics[width=11cm,angle=0]{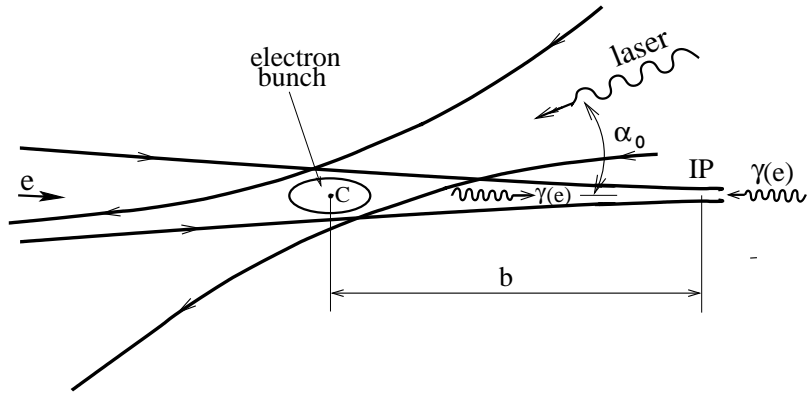}
\caption{A principal scheme of a photon collider. High-energy
electrons scatter on laser photons (in the conversion region C) and
produce high-energy $\gamma$ beam which collides with similar
$\gamma$ or $e$ beam at the interaction point IP.} \label{F3.2}
\end{figure}

The best solution for this task is to use a linear $e^{\pm}e^-$
collider as a basis and convert the $e^{\pm}$ beams into $\gamma$
beams by means of the backward Compton
scattering~\cite{GKST81a}---\cite{GKPST84}.

The principal scheme is shown in Fig.~\ref{F3.2}. An electron beam
after the final focus system is traveling towards the interaction
point IP. At the distance $b\, {\sim} 0.1 \div 1$ cm from the
interaction point, the electrons collide with the focused laser beam
in the conversion region C. The scattered high-energy photons follow
along the initial electron trajectories (with small additional
angular spread $\sim 1/\gamma_e$), hence,  {\bf the high-energy
photons are also focused at the interaction point IP}. This very
feature is one of the most attractive point in the discussed scheme.
The produced $\gamma $ beam collides downstream with the oncoming
electron or a similar $\gamma $ beam.~\footnote{To reduce
background, the ``used" electrons can be deflected from the
interaction point by an external magnetic field. In the scheme
without magnetic deflection background is somewhat higher, but such
scheme is simpler and allows to get higher luminosity.}

\begin{figure}[!htb]
\centering
\includegraphics[width=14cm,angle=0]{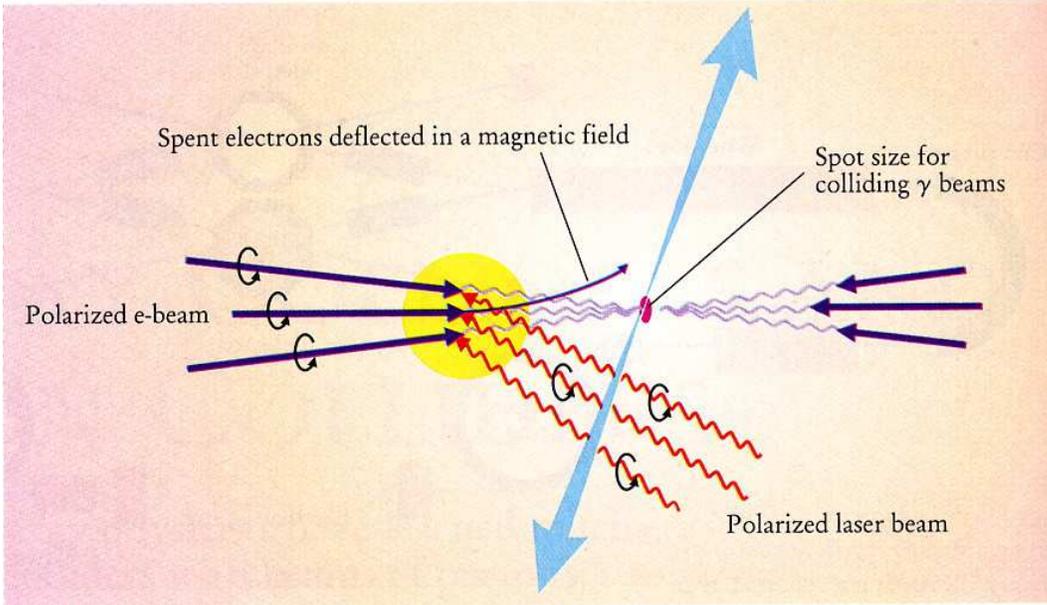}
\caption{Conversion region}
 \label{conversion_ses_2}
\end{figure}

More details about the conversion region are shown in Fig.
\ref{conversion_ses_2} from~\cite{Sessler}.

It is very important that modern laser technology allows to convert
most of electrons to high-energy photons. This means that the
$\gamma \gamma$ luminosity will be close to the luminosity of the
basic $e^{\pm}e^-$ beams.

\subsection{Compton scattering as a basic process for $e\to
\gamma$ conversion}

\begin{figure}[htb]
\begin{center}
\vspace{-0mm}
\hspace*{-1cm}\includegraphics[width=7cm,angle=0]{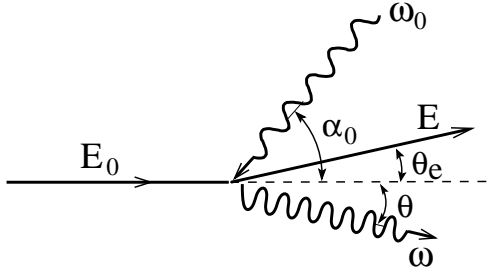}
\vspace{-0mm} \caption{Kinematics of the backward Compton
scattering} \label{F6.1} \vspace{-2mm}
\end{center}
\end{figure}

Properties of the linear and non-linear Compton scattering are
considered in details in~\cite{GKPST84}, \cite{KPS98}, \cite{IKS04}.
In the conversion region a laser photon with energy $\omega_{0} \sim
1$ eV scatters on an electron with energy $E_0 \sim 100$ GeV at a
small collision angle $\alpha _0$ (Fig.~\ref{F6.1}) and produces a
final photon with the energy $\omega$ and the emission angle
$\theta$. Kinematics of the backward Compton scattering
 \be
  e(p_0) +\gamma_0 (k_0) \to e(p) +\gamma (k)
 \label{6.1}
  \ee
is characterized by two dimensionless  variables $x$ and $y$:
\begin{equation}
x\;={2p_0 k_0\over m^2c^2}\approx {4E_0\omega_{0}\over m^{2}
c^4}\cos ^{2} {{\alpha_{0}\over 2}}\,, \;\;\; y\; = {kk_0\over
p_0k_0}\approx {\omega \over E_0}\,.
 \label{6.2}
\end{equation}
The maximum energy of the scattered photon $\omega_{m}$ and the
maximum value of the parameter $y$ are:
\begin{equation}
\omega \leq \omega_{m}\;= \; {x\over x+1}\, E_0\,, \;\;\; y\leq \;
y_m\; =\; {x\over x+1} = {\omega_{m}\over E_0}\;\,. \label{6.7}
\end{equation}
The energy of the scattered electron is $E= (1-y)E_0$ and its
minimum value is $E_{\min} = E_0/(x+1)$.

Typical example: in collision of the photon with $\omega_0 = 1.17$
eV ($\lambda =1.06\;\mu$m --- the region of the most powerful solid
state lasers) and the electron with $E_0= 250$ GeV, the parameter
$x=4.5$ and the maximum photon energy
  \be
\omega_m =0.82 E_0 = 205\;\;{\rm GeV}
 \ee
is close enough to the initial electron energy $E_0$.

A photon emission angle is very small, $\theta\sim 1/\gamma_e =
2\cdot 10^{-6}$.

\begin{figure}[htb]
\begin{center}
\vspace{-9mm}
\includegraphics[width=10cm,angle=0]{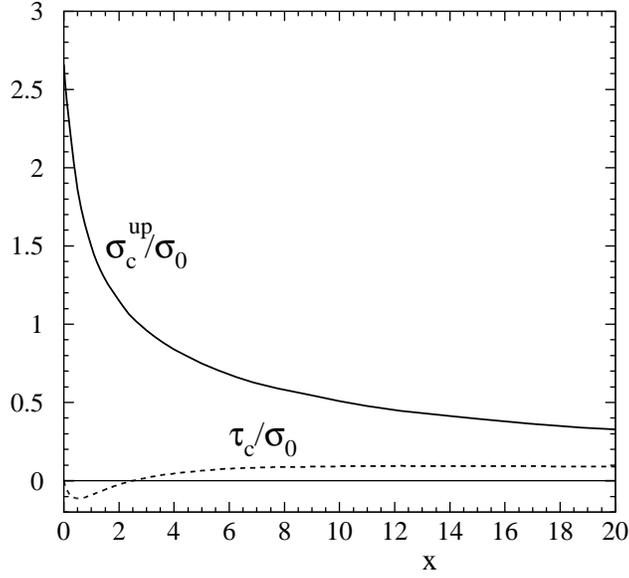}
\vspace{-8mm} \caption{Cross section $\sigma^{\mathrm{up}}_{{\mathrm
c}}$ for unpolarized photons and $\tau_{{\mathrm c}}$, related to
spin asymmetry (see (\ref{6.8}))}
 \label{F6.2}
\end{center}
\end{figure}

The total Compton cross section is
 $$
\sigma_{\mathrm c}\;=\;\sigma^{\mathrm{up}}_{\mathrm c}\;+
\;2\lambda_{e} P_{\mathrm c} \,\tau_{\mathrm c}\, ,
 $$
 where $\lambda_{e} $ is the mean helicity of the initial
electron, $P_c $ is that of the laser photon and
 \bea
\sigma^{\mathrm{up}}_{{\mathrm c}}&=&{2\sigma_{0}\over x}
\left[\left(1-{4\over x} - {8\over x^{2}}\right)\ln (x+1) +{1\over
2} +{8\over x} - {1\over 2(x+1)^{2}}\right] \, ,
 \label{6.8}
 \\
 \tau_{\mathrm c}&=& {2\sigma_{0}\over x} \left[\left(1+{2\over
x}\right) \ln (x+1)-{5\over 2}+{1\over x+1}-{1\over
2(x+1)^{2}}\right]\,,
  \nn\\
  \sigma_{0}&=& \pi r_e^2=2.5\cdot 10^{-25}\;\mbox {cm} ^{2}\,.
  \nn
 \eea
Note, that polarizations of initial beams influence the total cross
section (as well as the spectrum) only if both their helicities are
nonzero, i.e. at $\lambda_{e} P_{\mathrm c} \neq 0$. The
functions $\sigma^{\mathrm{up}}_{{\mathrm c}}$, corresponding to the
cross section of unpolarized beams, and $\tau_{{\mathrm c}}$,
determining the spin asymmetry, are shown in Fig.~\ref{F6.2}.
In the region of interest $x= 1 \div 5$ the  total  cross section is
large enough
 \begin{equation}
\sigma_{\mathrm c}\sim \sigma_0=2.5\cdot 10^{-25}\;\;{\rm cm}^2
 \label{6.12}
 \end{equation}
and only slightly depends  on  the  polarization of the initial
particles,  $|\tau_{\mathrm c}|/\sigma^{\mathrm{up}}_{\mathrm c}<
0.1$.

On the contrary, the energy spectrum does essentially depend on the
value of $\lambda_{e} P_{\mathrm c}$. The energy spectrum of
scattered photons is defined by the differential Compton cross
section:
 \be
{d\sigma _{{\mathrm c}}\over dy} ={2\sigma_{0}\over x} \left[{1\over
1-y} + 1-y - 4r(1-r) - 2\lambda_{e} P_c\;{y(2-y)\over
1-y}\,(2r-1)\right]\,.
 \label{6.13}
 \ee
It is useful to note that $r=y/[x(1-y)] \leq 1$ and $r\to 1$ at
$y\to y_m$.
\begin{figure}[!h]
\begin{center}
\vspace{-8mm}
\includegraphics[width=10cm,angle=0]{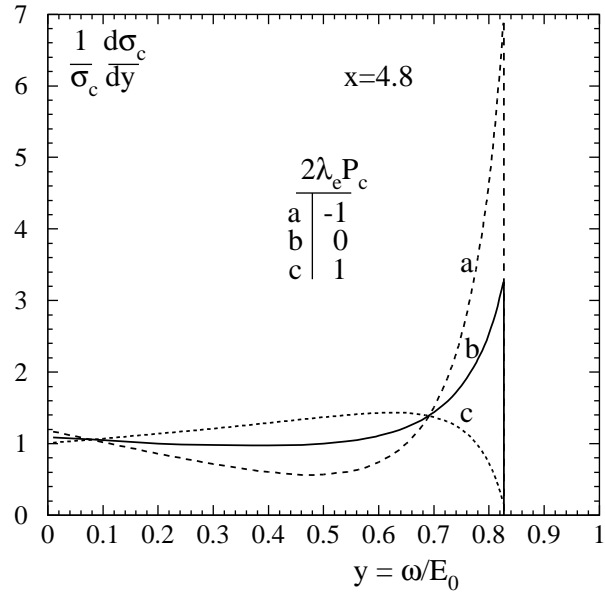}
\vspace{-8mm} \caption{Energy spectrum of scattering photons at
$x=4.8$} \label{F6.3}
\end{center}
\end{figure}

\begin{figure}[!h]
\begin{center}
\vspace{-5mm}
\hspace*{-0.5cm}\includegraphics[width=7cm,angle=0]{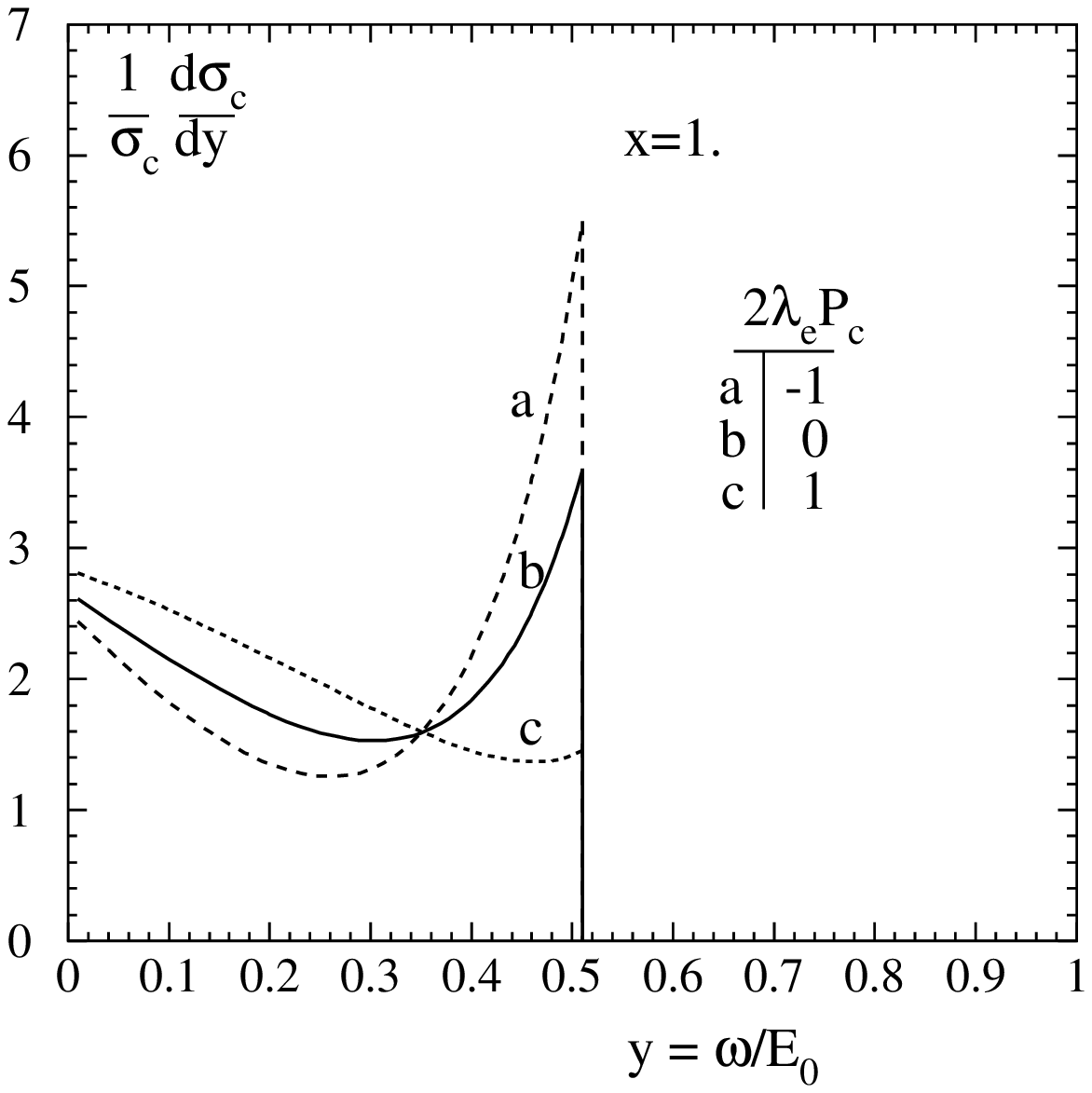}\hspace*{-0.7cm}
\includegraphics[width=7cm,angle=0]{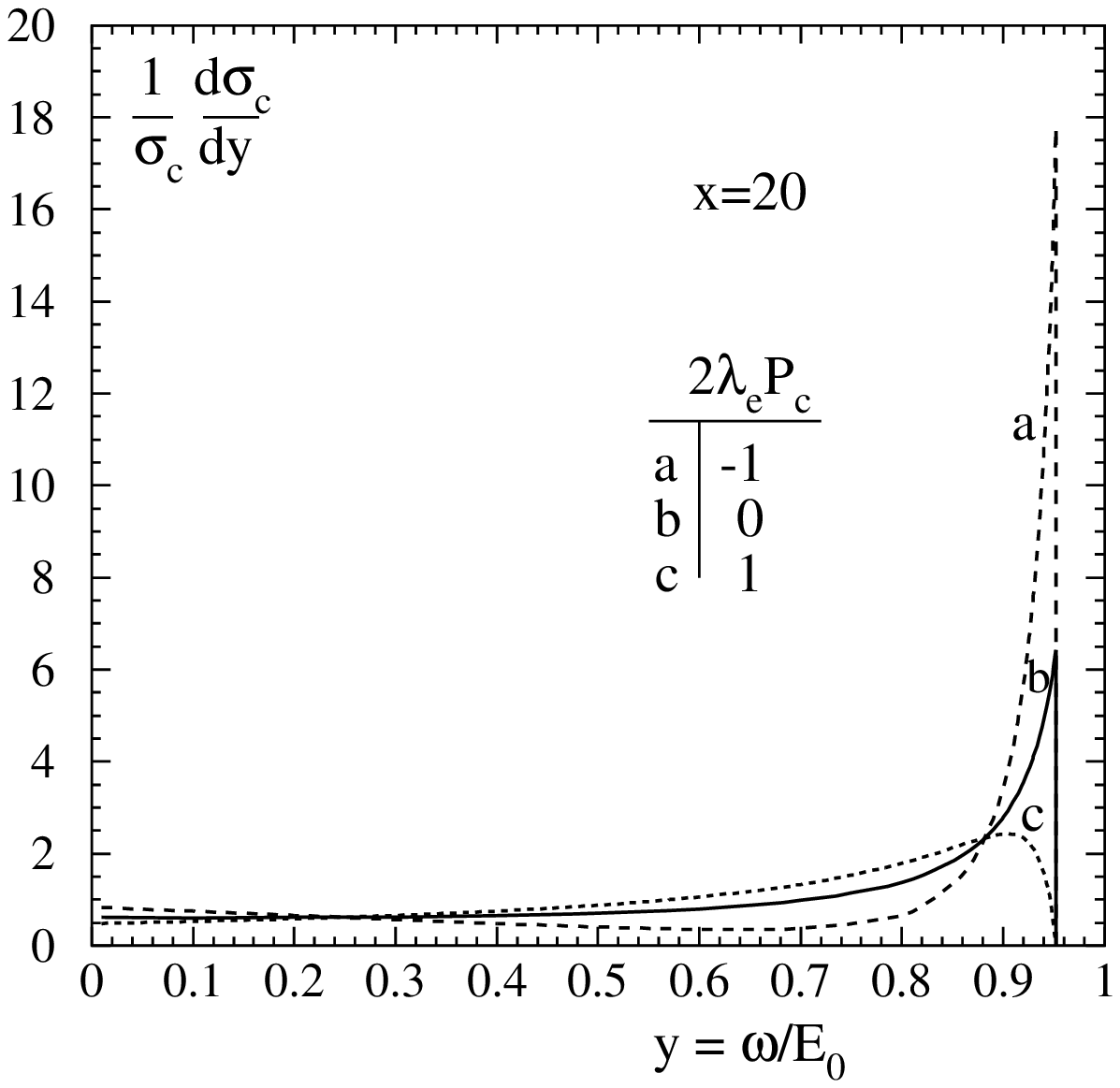}
\vspace{-5mm}
 \caption{Energy spectrum of scattering photons at
$x=1$ (left figure) and $x=20$ (right figure)}
 \label{F6.4}
 \vspace{-2mm}
\end{center}
\end{figure}
The ``quality'' of the photon beam, i.e. the relative number of hard
photons, is better for the negative value of $\lambda_{e}
P_{\mathrm c}$. For $2\lambda_{e} P_c=-1$ the peak value of the
spectrum at $\omega =\omega_{m}$ nearly doubles improving
significantly the monochromaticity of the $\gamma$ beam (cf. curves
$a$ and $b$ in Fig.~\ref{F6.3}).

In order to increase the maximum photon energy, one should use the
laser with larger frequency. This also increases a fraction of hard
photons (cf. Figs.~\ref{F6.4}  and \ref{F6.3}). Unfortunately, at
large $x$ the high energy photons disappear from the beam producing
$e^{+} e^{-}$ pairs in collisions with laser photons. The threshold
of this reaction, $\gamma\gamma_L\to e^+e^-$, corresponds to $x
\approx 4.8$. Therefore, it seems that the value $x\approx 5$ is the
most preferable.

\subsubsection{Angular distribution}

The energy of a scattered photon depends on its emission angle
$\theta $ as follows:
 \begin {equation}
\omega ={\omega _m \over 1+(\theta /\theta_0)^2}; \quad \theta_{0}
\; = \; {m c^2 \over E_0}\; \sqrt{x+1}\,.
 \label{6.16}
 \end{equation}
Note, that photons with the maximum energy $\omega_m$ scatter at
zero angle. The angular distribution of scattered protons has a very
sharp peak in the  direction of the incident electron momentum.

After the Compton scattering, both electrons and photons travel
essentially along the original electron beam direction. The photon
and electron scattering angles ($\theta$ and $\theta_e$) are unique
function of the photon energy:
 \begin {equation}
\theta (y)=\theta_0\; \sqrt{{y_m\over y}-1}, \;\;\; \;
\theta_e={\theta_0\; \sqrt{y(y_m-y)} \over 1-y}\; .
 \label{6.21}
 \end {equation}
\begin{figure}[!h]
\begin{center}
\vspace{-4mm}
\includegraphics[width=8cm,angle=0]{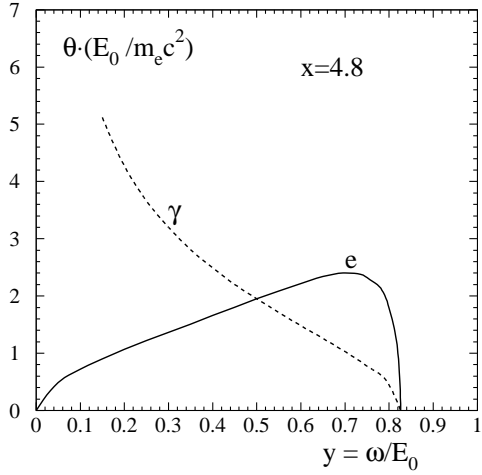}
\vspace{-6mm} \caption{ Photon and electron scattering angles vs the
photon energy $\omega$  at $x=4.8$} \label{F6.5} \vspace{-6mm}
\end{center}
\end{figure}
For $x=4.8$ these  functions are plotted in Fig.~\ref{F6.5}. It is
remarkable that electrons are only slightly deflected from their
original direction and scatter into a narrow cone:
\begin {equation}
\theta_e \leq {x \over 2\gamma} = {2\omega _0 \over m c^2} .
\label{6.22}
\end {equation}

\subsubsection{Polarization of final photons}

Using the polarized initial electrons and laser photons, one can
obtain the high-energy photons with various polarization. Let us
present two typical examples.

\begin{figure}[!h]
\begin{center}
\vspace{-8mm}
\includegraphics[width=9cm,angle=0]{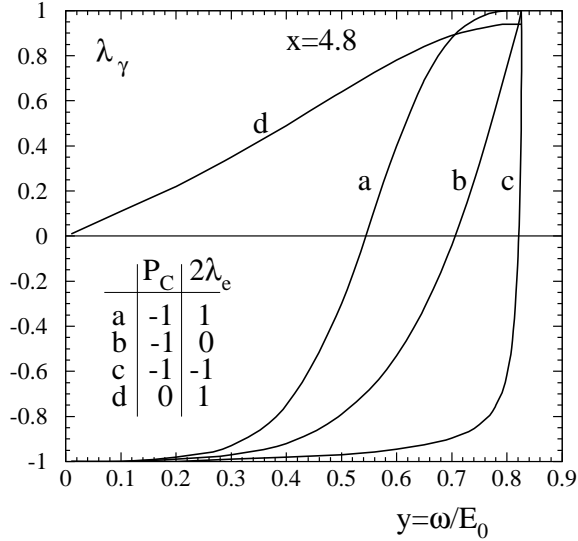}
\vspace{-5mm} \caption{Mean helicity of the scattered photons
$\lambda_\gamma$ vs. $\omega /E_0$ for various laser photon
helicities $P_c$ and electron helicities $\lambda _e$ at $x= 4.8$ }
\label{F6.6} \vspace{-2mm}
\end{center}
\end{figure}
The mean helicity of the final photon $\lambda_\gamma$ in dependence
on the final photon energy $\omega$ is shown in Figs.~\ref{F6.6}.
Curves $a,\; b$ and $c$ in Fig.~\ref{F6.6} correspond to spectra
$a,\; b$ and $c$ in Fig.~\ref{F6.3}. In the case $2\lambda_e
P_c = -1$ (the case of good monochromaticity -- see curves $a$ in
Fig.~\ref{F6.3}) almost all high-energy photons have a high degree
of circular polarization. In the case $b$ the electrons are
unpolarized, and the region, where high-energy photons have a high
degree of polarization, is much narrow.

\begin{figure}[!h]
\begin{center}
\vspace{-8mm}
\includegraphics[width=9cm,angle=0]{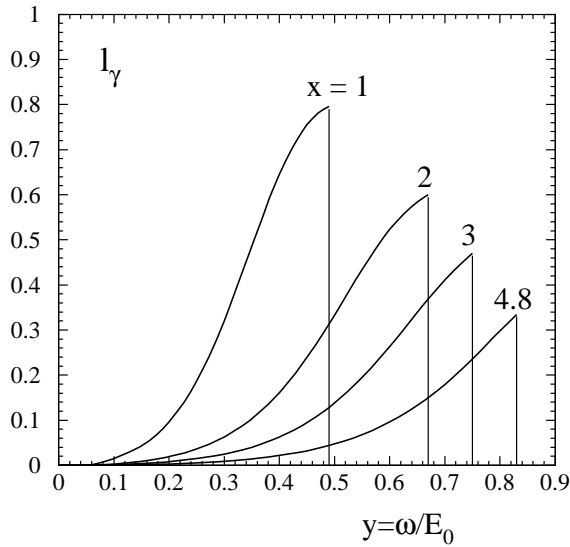}
\vspace{-5mm}
 \caption{Average linear polarization of the scattered
photons $l_\gamma$  vs. $\omega/E_0$ at $x=1,\; 2,\; 3$ and $4.8$
(the degree of linear polarization of the laser photon $P_l = 1$)}
 \label{F6.8}
\end{center}
\end{figure}
If the  laser light has a linear polarization, then high-energy
photons are polarized in the same direction. The average degree of
the linear polarization of the final photon $l_\gamma$ in dependence
on the final photon energy $\omega$ is shown in Figs.~\ref{F6.8}.
The linear polarization of the colliding photons is very important
for study of the nature of Higgs boson.

\section{Physics of $\gamma\gamma$ Interactions }

\begin{figure}[!ht]
\centering
\includegraphics[width=14cm,angle=0]{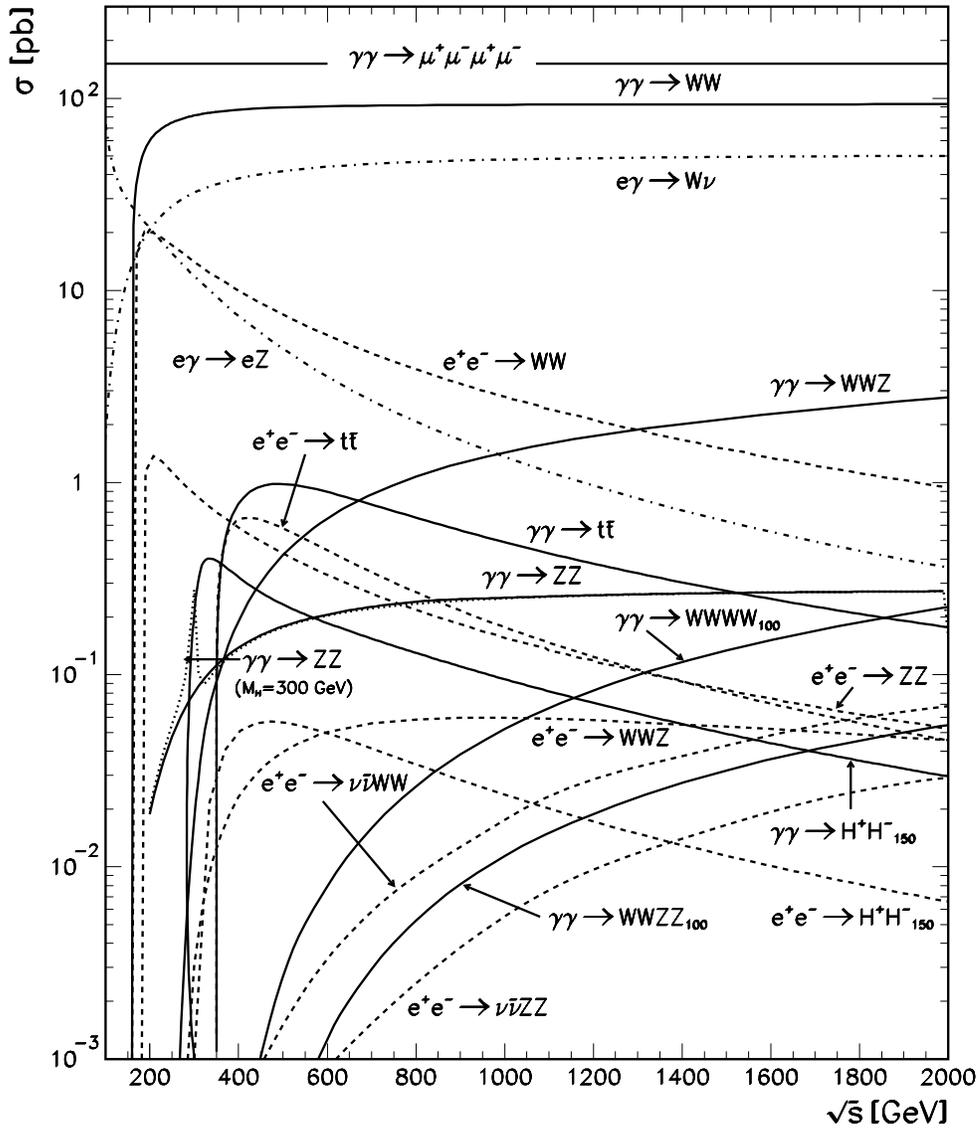}
 \caption{Cross sections for some interesting processes in
 $e^+e^-$, $\gamma e$ and $\gamma \gamma$ collisions}
 \label{all-cross}
 \end{figure}

Physical potential of such $\gamma \gamma$ and $\gamma e$ colliders
will be on the same level with future $e^+e^-$ and $pp$ colliders.
Moreover, there is a number of problems in which photon colliders
are beyond competition. The comparison of cross sections in the
$e^+e^-$, $\gamma \gamma$ and $\gamma e$ colliders are given in
Fig.~\ref{all-cross}.

Photon collider (PC) makes it possible to investigate both problems
of new physics and of ``classical" hadron physics and QCD.

Since photon couple directly to all fundamental charged particles
--- leptons, quarks, $W$ bosons, super-symmetric particles, etc.
--- a PC can provide a possibility to test every aspect of the
Standard Model (SM) and beyond.

\begin{figure}[!ht]
 \centering
 \includegraphics[width=7cm,angle=0]{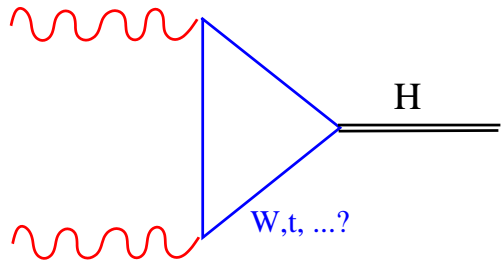}
  \caption{Higgs boson production in $\gamma \gamma$ collision}
   \label{gg-to-higgs}
  \end{figure}

Besides, photons can couple to neutral particles (gluons, $Z$
bosons,  Higgs bosons, etc.) through charged particles box diagrams
(Fig. \ref{gg-to-higgs}).

On the other hand, in a number of aspects photons are similar to
hadrons, but with simpler initial state. Therefor, PC will be
perfect in studying of QCD and other problems of hadron physics.

Let us list the problems in which the photon colliders have a high
potential or some advantages:

{\it Higgs hunting and investigation}. PC provides the opportunity
to observe the Higgs boson at the smallest energy as a resonance in
the $\gamma \gamma$ system, moreover, PC is out of competition in
the testing of Higgs nature.

\begin{figure}[!t]
\centering
\includegraphics[width=15cm,angle=0]{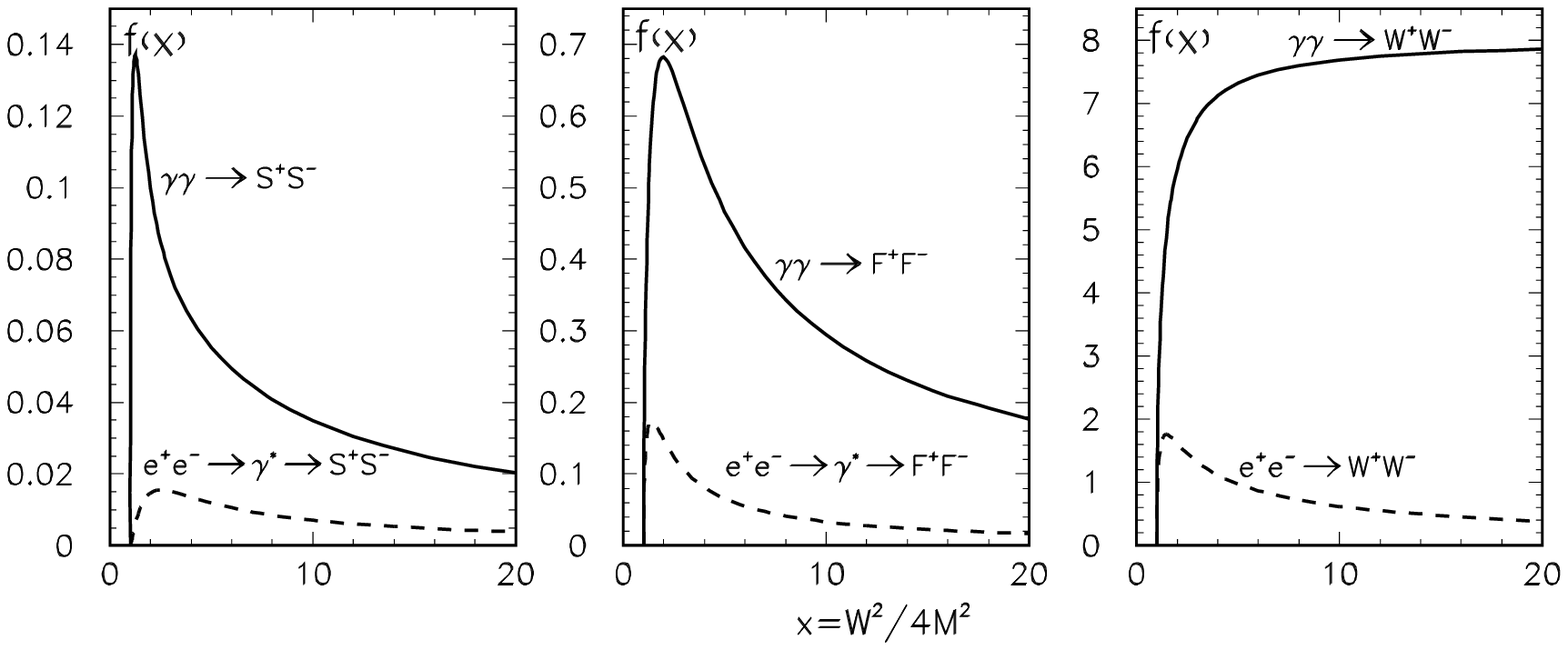}
 \caption{Cross sections for charged pair production in
 $e^+e^-$ and $\gamma \gamma$ collisions: here $S$ -- scalars, $F$ -- fermions,
$\sigma= (\pi \alpha^2/M^2)\, f(x)$}
 \label{ggtoSS}
\end{figure}

{\it Beyond SM}. PC provides the excellent opportunities for finding
of various particles beyond the SM: SUSY partners, charged Higgses,
excited leptons and quarks, leptoquarks,... In particular, $\gamma
e$ collider will be the best machines for discovery of selectron or
excited electron. Cross sections for the charged pair production in
$\gamma \gamma$ collisions are larger than in $e^+e^-$ collisions
--- see Fig. \ref{ggtoSS}.

{\it Electroweak gauge boson physics}. The electroweak theory is the
substantial part of the SM which pretend for precise description
like QED. PC will be $W$ factories with a rate about $10^7\; W$
bosons per year. In addition, the $\gamma e \to W \nu$ process will
produce  single $W$s which is very attractive for $W$ decay's study.
Thus, PCs provide one of the best opportunity to test the precise
predictions of the electroweak theory.

{\it QCD and hadron physics}. The photon colliders provide the
unique possibility to investigate the problems of hadron physics and
QCD in the new type of collisions and with the simplest structure of
initial state. The principal topics here are the following:
\begin{itemize}
\item{the $t \bar t $ production in different partial waves;}
\item{the photon structure functions;}
\item{the semihard processes;}
\item{the jet production;}
\item{the total $\gamma \gamma \to hadrons$ cross section.}
\end{itemize}

To clarify many of these points it will be very useful  to compare
results from $\gamma \gamma,\; \gamma e,\; ep$ and $pp$ colliders.

Besides the high-energy $\gamma \gamma $ and $\gamma e$ collisions,
PCs provide {\it some additional options}:

{\it (i)} The region of conversion $e\to \gamma$ can be treated as
$e\gamma_L$ collider (here $\gamma _L$ is the laser photon) with
c.m.s. energy $\sim 1 $ MeV but with enormous luminosity $\sim
10^{38}\div10^{39}$ cm$^{-2}$s$^{-1}$.  It can be used, for example,
for search of weakly interacting light particles, like invisible
axion.

{\it (ii)} In the conversion region one can test non-linear QED
processes, like the $e^+e^-$ pair production in collision of
high-energy photon with a few laser photons.

{\it (iii)} The used high-energy photon beams can be utilized for
fixed-target experiments.

\section{Concluding remarks}

\subsection{Summary from TESLA TDR}

The TESLA Technical Design Report~\cite{TESLA} contains the part
devoted to the high-energy photon collider~\cite{TESLA-04}. It will
be useful to cite the  Physics Summary from this part:

 \begin{quotation}

To summarize, the Photon Collider will allow us to study the physics
of the electroweak symmetry breaking in both the weak-coupling and
strong-coupling scenario.

 Measurements of the two-photon Higgs width of the $h,\,H$ and $A$
Higgs states provide a strong physics motivation for developing the
technology of the $\gamma \gamma$ collider option.

Polarized photon beams, large cross sections and sufficiently large
luminosities allow to significantly enhance the discovery limits of
many new particles in SUSY and other extensions of the Standard
Model.

Moreover, they will substantially improve the accuracy of the
precision measurements of anomalous W boson and top quark couplings,
thereby complementing and improving the measurements at the $e^+e^-$
mode of TESLA.

Photon colliders offer a unique possibility for probing the photon
structure and the QCD Pomeron.

 \end{quotation}

\subsection{Prediction of Andrew Sessler~\cite{Sessler} in 1998 }

\begin{quotation}

At present, Europe has the lead in electron colliders (LEP), hadron
colliders (LHC) and hadron-electron colliders(HERA). Stanford and
Japan's High Energy Research Organization (KEK) are jointly working
on a TeV $e^+e^-$ collider disign, as in DESY. Japan and/or Germany
seem to be the most likely location for the next-generation $e^+e^-$
machine. Looking broadly, and also contemplating what US will do in
high-energy physics, one may imagine a $\mu^+\mu^-$ collider in the
US, early in the next century.

\end{quotation}

\subsection{Conclusion of Karl von Weizs\"acker for young physicists}

I would like to tell you a little real story. In 1991 the First
International Conference on Physics devoted to Andrej Sakharov held
in Moscow. A number of great man have participated in this
Conference including a dozen of Nobel prize winners. Among others
was Karl von Weizs\"acker.

I personally was very interesting in application of the
Weizs\"acker-Williams method to the two-photon processes at
colliding beams and I even have published a few articles on this
subject. So, I would like to see and to speak with the author of
this method. At the beginning of our conversation, Weizs\"acker told
me a history of this invention.

In 1934 Weizs\"acker was in Copenhagen as an assistant of Prof.
Niels Bohr. And just at that moment there was some international
Conference on Physics in the N.~Bohr Institute. Williams was the
first who suggested the idea of the equivalent photon approximation
for QED processes. But the final result appeared only after a lot of
heat discussions in which Weizs\"acker, Williams, Niels Bohr, Lev
Landau, Edward Teller and some others participated.

After the Conference all people, but Weizs\"acker, went back to
their home institutes. So, it was quite natural that some day
N.~Bohr invited  Weizs\"acker and said: ``And you, young man, you
should write a paper on the discussed subject''.

And  Weizs\"acker did and brought the paper to Bohr.  A few days
after that Bohr invited him to discuss the work. ``At the beginning
of this meeting, --- Weizs\"acker told me, --- I was young and
exited, but Professor seemed to me was old and tired''.

Bohr said:``It is an excelent work, a very clean and perfect paper,
but... I have a small remark about the second page''. So they have
discussed this small remark. After that they have discussed another
remark, and another remark, and another...

``After four or five hours of discussion. ---  Weizs\"acker
continued, --- I was young and tired, but Professor was old and
excited''. At the end Bohr said:``Now I see that you wrote quite
contrary to what you think, you should rewrite your paper''.

 Weizs\"acker's conclusion was: ``I think that it is the best way for a
 young scientist to study physics: you should have a good problem
 and a possibility to discuss this problem during hours with a great man''.


\end{document}